\documentclass[twocolumn,showpacs,prb,preprintnumbers,amsmath,amssymb]{revtex4}

\usepackage{graphicx}
\usepackage{dcolumn}
\usepackage{bm}
\usepackage{epsfig}

\begin{document}


\title{
Effect of magnetic pair breaking on Andreev bound states and 
resonant supercurrent in quantum dot Josephson junctions
}

\author{Grygoriy Tkachov$^{1,2}$ and Klaus Richter$^{1}$}
\affiliation{
$^{1}$ Institute for Theoretical Physics, Regensburg University, 93040 Regensburg, Germany\\
$^{2}$ Max Planck Institute for the Physics of Complex Systems, 01187 Dresden, Germany
}


\begin{abstract}
We propose a model for resonant Josephson tunneling through quantum dots that  
accounts for Cooper pair-breaking processes in the superconducting leads 
caused by a magnetic field or spin-flip scattering.  
The pair-breaking effect on the critical supercurrent $I_c$ and the Josephson current-phase 
relation $I(\varphi)$ is largely due to the modification of the spectrum of Andreev bound states 
below the reduced (Abrikosov-Gorkov) quasiparticle gap.
For a quantum dot formed in a quasi-one-dimensional channel, 
both $I_c$ and $I(\varphi)$ can show a significant magnetic field dependence
induced by pair breaking 
despite the suppression of the orbital magnetic field effect in the channel. 
This case is relevant to recent experiments on quantum dot Josephson junctions in 
carbon nanotubes.
Pair-breaking processes are taken into account via 
the relation between the Andreev scattering matrix and 
the quasiclassical Green functions of the superconductors in the Usadel limit.
\end{abstract}


\maketitle
\section{Introduction}
 
Since its discovery the Josephson effect~\cite{Josephson} 
has been studied for a variety of 
superconducting weak links~\cite{Likharev,Tinkham,Golubov}. 
The research has recently entered a new phase with 
the experimental realization of quantum dot weak links
exploiting electronic properties of finite-length carbon nanotubes 
coupled to superconducting leads~\cite{Basel,Jarillo,Jorgensen}.
In particular, for the first time 
since its theoretical prediction~\cite{Matveev,vanHouten,Beenakker}
resonant Josephson tunneling through discrete electronic states 
has been observed in carbon nanotube quantum dots~\cite{Jarillo}. 
As demonstrated in Refs.~\onlinecite{Jarillo,Jorgensen},
the novel type of weak links exhibits transistor-like 
functionalities, 
e.g. a periodic modulation of the critical current with a gate voltage 
tuning successive energy levels 
in the dot on- and off-resonance with the Fermi energy in the leads.  
This property has already been implemented in a recently 
proposed carbon nanotube superconducting quantum interference 
device (CNT-SQUID)~\cite{SQUID} with possible applications 
in the field of molecular magnetism.  

Motivated by the experiments on resonant Josephson tunneling, 
in this paper we investigate theoretically 
how robust it is with respect to pair-breaking perturbations
in the superconducting leads.
Cooper pair breaking can be induced by a number of factors, 
e.g. by paramagnetic impurities~\cite{AG}, 
an external magnetic field~\cite{Maki} 
or by structural inhomogeneities producing
spatial fluctuations of the superconducting coupling constant~\cite{Fluctuations}.
It can cause a drastic distortion of 
the Bardeen-Cooper-Schrieffer (BCS) superconducting 
state, which manifests itself in the smearing of the BSC density of states 
leading to gapless superconductivity~\cite{AG,Maki}. 

While the pair-breaking effect on bulk superconductivity is now 
well understood, its implications for quantum superconducting transport 
have been studied to a much lesser extent 
[see, e.g. Refs.~\onlinecite{Golubov, Averin, Cuevas}]
which to our knowledge does not cover 
Josephson tunneling through quantum dots.  
On the other hand, in low-dimensional systems pair-breaking effects 
may be observable in a common experimental situation when,  
for instance, a carbon nanotube weak link is subject to a magnetic field. 
Since the orbital field effect in the quasi-one-dimensional 
channel is strongly suppressed, 
pair breaking in the superconducting leads 
can be the main source of the magnetic field dependence 
of the Josephson current. This situation is addressed in our work.   

\begin{figure}[b]
\epsfxsize=0.5\hsize
\epsffile{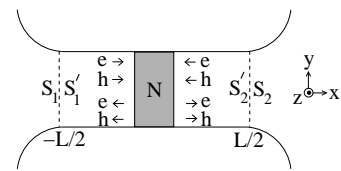}
\caption{Scheme of a superconducting constriction 
with a normal scattering region $N$. 
The arrows indicate the electrons (e) and holes (h) 
incident on and outgoing from $N$.}
\label{Cons}
\end{figure}
%

The influence of pair breaking on the Josephson current can not, in general, 
be accounted for by mere suppression of the order parameter in the superconducting leads.
As was pointed out in Ref.~\onlinecite{Averin}, it is a more subtle effect involving 
the modification of the spectrum of current carrying states in the junction, in particular, 
the subgap states usually referred to as Andreev bound states (ABS)~\cite{Beenakker,BS}.
We illustrate this idea for quantum dot junctions in the simple model  
of a short superconducting constriction 
with a scattering region containing a single Breit-Wigner resonance 
near the Fermi energy.   
The Josephson current is calculated using
the normal-state scattering matrix of the system 
and the Andreev reflection matrix~\cite{vanHouten,Beenakker}. 
Unlike Refs.~\onlinecite{vanHouten,Beenakker}
we focus on dirty superconductors 
for which the Andreev matrix can be quite generally expressed in terms 
of the quasiclassical Green functions~\cite{GolKup},
allowing us to treat pair breaking in the superconducting leads nonperturbatively. 
Although we account for all energies 
(below and above the Abrikosov-Gorkov gap $\Delta_{\bf g}$),
it turns out that the behavior of the Josephson current can be well understood 
in terms of a pair-breaking-induced modification of the ABS, 
which depends sensitively on the relation between the Breit-Wigner 
resonance width $\Gamma$ and the superconducting pairing energy $\Delta$.
Both the critical supercurrent and the Josephson current-phase relation 
are analyzed under experimentally realizable conditions.

\section{Model and Formalism}

We consider a junction between two  
superconductors $S_1$ and $S_2$ 
adiabatically narrowing into quasi-one-dimensional ballistic wires 
$S^\prime_1$ and $S^\prime_2$ coupled to
a normal conductor $N$ [Fig.~\ref{Cons}]. 
The transformation from the superconducting electron spectrum to 
the normal-metal one is assumed to take place  
at the boundaries $S_1S^\prime_1$ and $S^\prime_2S_2$,
implying the pairing potential of the form~\cite{Likharev}: 
$\Delta(x)=\Delta{\rm e}^{i\varphi_1}$ for $x<-L/2$, 
$\Delta(x)=0$ for $|x|\leq L/2$ 
and $\Delta(x)=\Delta{\rm e}^{i\varphi_2}$ for $x>L/2$
with the order parameter phase difference $\varphi\equiv\varphi_2-\varphi_1$ 
and the junction length $L\ll\hbar v_F/\Delta$ 
($v_F$ is the Fermi velocity in $S_{1,2}$).

The Josephson coupling can be interpreted in terms of the Andreev process~\cite{Andreev} 
whereby an electron is retro-reflected as a Fermi-sea hole 
from one of the superconductors with the subsequent hole-to-electron 
conversion in the other one. 
Such an Andreev reflection circle facilitates a Cooper pair transfer 
between $S_1$ and $S_2$. 
Normal backscattering from disordered superconducting bulk 
into a single-channel junction is suppressed 
due to the smallness of the junction width compared to 
the elastic mean free path $\ell$.
The $N$ region in the middle of the junction 
is thus supposed to be the only source of normal scattering. 
In such type of weak links the Josephson current
is conveniently described by the scattering matrix expression 
of Refs.~\onlinecite{Beenakker,Brouwer} 
that can be written at finite temperature $T$ 
as the following sum over the Matsubara frequencies 
$\omega_n=(2n+1)\pi k_BT$~[Ref.~\onlinecite{Brouwer}]:
\!
\begin{eqnarray}
I=-\frac{2e}{\hbar}\, 2k_BT\,\frac{\partial}{\partial\varphi}
\sum\limits_{n=0}^{\infty}
\ln\,{\rm Det}\left[
\hat{1}-\hat{s}_A(E)\hat{s}_N(E)
\right]_{E=i\omega_n.}
\label{I}
\end{eqnarray} 
Here $\hat{s}_N(E)$ is a $4\times 4$ unitary matrix 
relating the incident electron and hole waves on the $N$ region 
to the outgoing ones [Fig.~\ref{Cons}]. 
It is diagonal in the electron-hole space:

\begin{eqnarray}
\hat{s}_N=
\left[
\begin{array}{cc}
s_{ee}(E) & 0\\
0 & s_{hh}(E)
\end{array}
\right],
\,
s_{ee}(E)=
\left[
\begin{array}{cc}
r_{11}(E) & t_{12}(E)\\
t_{21}(E) & r_{22}(E)
\end{array}
\right].
\nonumber
\end{eqnarray}
The matrix $s_{ee}(E)$ describes electron
scattering in terms of the reflection  
and transmission amplitudes,  $r_{jk}(E)$  and $t_{jk}(E)$,
for a transition from $S^\prime_k$ to $S^\prime_j$ 
($j,k=1,2$). 
The hole scattering matrix is related 
to the electron one by $s_{hh}(E)=s^*_{ee}(-E)$.
The Andreev scattering matrix $\hat{s}_A(E)$ 
is off-diagonal in the electron-hole space:

\begin{eqnarray}
\hat{s}_A=
\left[
\begin{array}{cc}
0 & s_{eh}(E)\\
s_{he}(E) & 0
\end{array}
\right],
\label{A}
\end{eqnarray}
where the $2\times 2$ matrices $s_{he}(E)$ and $s_{eh}(E)$ 
govern the electron-to-hole and hole-to-electron scattering 
off the superconductors. 
Equation (\ref{I}) is valid for all energies 
as long as normal scattering from the superconductors is absent~\cite{Beenakker,Brouwer}. 

In Ref.~\onlinecite{Beenakker} the Andreev matrix (\ref{A})  
was obtained by matching 
the solutions of the Bogolubov-de Gennes 
equations in the wires $S^\prime_{1,2}$
to the corresponding solutions 
in impurity-free leads. 
Gorkov's Green function formalism
in combination with the quasiclassical theory~\cite{Quasi}
allows one to generalize the results of Ref.~\onlinecite{Beenakker} 
to dirty leads with a short mean free path 
$\ell\ll\hbar v_F/\Delta$.
In the latter case the matrices $s_{he}(E)$ and $s_{eh}(E)$
can be expressed in terms of the 
quasiclassical Green functions of the superconductors
as follows~\cite{GolKup}:

\begin{eqnarray*}
s_{eh}=
\left[
\begin{array}{cc}
\frac{f_1(E)}{g_1(E)+1} & 0\\
0 & \frac{f_2(E)}{g_2(E)+1}
\end{array}
\right],
\,
s_{he}=
\left[
\begin{array}{cc}
\frac{-f^\dagger_1(E)}{g_1(E)+1} & 0\\
0 & \frac{-f^\dagger_2(E)}{g_2(E)+1}
\end{array}
\right].
\end{eqnarray*}
Here $g_{1,2}$ and $f_{1,2}$ ($f^\dagger_{1,2}$) 
are, respectively, the normal and anomalous retarded Green 
functions in $S_{1,2}$. 
These matrices are diagonal in the electrode space 
due to a local character of Andreev reflection in our geometry.

Neglecting the influence of the narrow weak link 
on the bulk superconductivity, 
we can use the Green functions of 
the uncoupled superconductors $S_{1,2}$ 
described by the position-independent 
Usadel equation~\cite{Quasi},
\,
\begin{eqnarray}
\left[E\hat\tau_3+\hat\Delta_j+\frac{i\hbar}{2\tau_{pb}}\,
\hat\tau_3\hat g_j\hat\tau_3\,,\,\hat g_j\right]=0,
\label{Usadel}
\end{eqnarray}
with the normalization condition $\hat g_j^2=\hat\tau_0$
for the matrix Green function  
\,
\begin{eqnarray*}
\hat g_j=\left[
\begin{array}{cc}
g_j & f_j\\
f^\dagger_j & -g_j
\end{array}
\right],
\,\,
\hat\Delta_j=\left[
\begin{array}{cc}
0 & \Delta {\rm e}^{i\varphi_j}\\
-\Delta {\rm e}^{-i\varphi_j} & 0
\end{array}
\right], 
\,\,
j=1,2.	
\end{eqnarray*}
Here $\hat\tau_0$ and $\hat\tau_3$ are the unity and Pauli matrices, 
respectively, and $[...,...]$ denotes a commutator. 
Equation~(\ref{Usadel}) accounts for a finite pair-breaking 
rate $\tau^{-1}_{pb}$ whose
microscopic expression depends on the nature of 
the pair-breaking mechanism.
For instance, for thin superconducting films in a parallel magnetic field, 
$\tau^{-1}_{pb}=(v_F\ell/18)(\pi dB/\Phi_0)^2$~[Ref.~\onlinecite{Maki}] 
where $d$ is the film thickness and $\Phi_0$ is the flux quantum.
For paramagnetic impurities, 
$\tau_{pb}$ coincides with the spin-flip time~\cite{AG}.
In the case of the spatial fluctuations of the superconducting coupling, 
$\tau^{-1}_{pb}$ is proportional to the variance of the fluctuations~\cite{Fluctuations}. 

From Eq.~(\ref{Usadel}) one obtains the Green functions
\!
\begin{eqnarray}
&&
g_j=\frac{u}{\sqrt{u^2-1}}=
u{\rm e}^{-i\varphi_j} f_j,\quad 
f^\dagger_j=-{\rm e}^{-2i\varphi_j}f_j,
\label{Green}\\
&&
\frac{E}{\Delta}=u\left(1-\frac{\zeta}{\sqrt{1-u^2}}\right),\quad
\zeta=\frac{\hbar}{\tau_{pb}\Delta},
\label{u}
\end{eqnarray}
where, following Refs.~\onlinecite{AG,Maki}, 
we introduce a dimensionless pair-breaking parameter $\zeta$. 
The matrices $s_{eh}$ and $s_{he}$ 
can be expressed using Eqs.~(\ref{Green}) as follows:
\!
\begin{eqnarray}
&
s_{eh}=\alpha 
\left[
\begin{array}{cc}
{\rm e}^{i\varphi_1} & 0\\
0 & {\rm e}^{i\varphi_2}
\end{array}
\right],
\,\,
s_{he}=\alpha
\left[
\begin{array}{cc}
{\rm e}^{-i\varphi_1} & 0\\
0 & {\rm e}^{-i\varphi_2}
\end{array}
\right],&
\label{SA}\\
&\alpha=u-\sqrt{u^2-1}.&
\label{a}
\end{eqnarray}
We note that pair breaking modifies 
the energy dependence of the Andreev reflection amplitude 
$\alpha$ according to the non-BCS Green functions (\ref{Green}) and (\ref{u}). 
A few words concerning the applicability of this result are due here.

First of all, there is no restriction on energy $E$, e.g. 
for $\zeta\leq 1$, equations (\ref{SA}) and (\ref{a}) are 
valid both below and above the reduced (Abrikosov-Gorkov) 
quasiparticle gap $\Delta_{\bf g}=\Delta\left(1-\zeta^{2/3}\right)^{3/2}$. 
In particular, for $|E|\leq\Delta_{\bf g}$ one can show that 
$u$ is real and $|u|\leq(1-\zeta^{2/3})^{1/2}<1$ [Ref.~\onlinecite{Maki}], 
corresponding to perfect Andreev reflection with  
$\alpha=\exp(-i\arccos(u))$. 
Since in the Usadel limit $\ell\ll v_F\tau_{pb}$, 
normal scattering from the superconductors is suppressed 
due to the smallness of the junction width 
also in the presence of pair breaking. 
The absence of normal transmission at $|E|\leq\Delta_{\bf g}$
is consistent with the Abrikosov-Gorkov approach
assuming no impurity states inside the gap and  
the validity of the Born approximation~\cite{AG,Maki}. 
For $|E|\geq\Delta_{\bf g}$ the relevant solution of Eq.~(\ref{u}) 
is complex and has positive ${\rm Im}u$ related to the density of states 
of the superconductor~\cite{Maki}.  
Equations (\ref{u}) and (\ref{a}) are thus the generalization 
of the known result $\alpha=(E/\Delta_0)-\sqrt{(E/\Delta_0)^2-1}$ [Ref.~\onlinecite{BTK}]
for transparent point contact, 
where $\Delta_0\equiv\Delta|_{\zeta=0}$ is the BCS gap.  
It is convenient to measure all energies in units of $\Delta_0$ 
for which equations (\ref{Green})--(\ref{a}) should be complemented with 
the self-consistency equation for $\Delta$. 
At $T=0$, the case we are eventually interested in, this equation can be written as
~\cite{AG,Maki}:
\!
\begin{eqnarray}
\ln(\zeta_0/\zeta)&=&-\pi\zeta/4, \qquad\zeta\leq 1,
\label{OP1}\\
\ln(\zeta_0/\zeta)&=&\sqrt{\zeta^2-1}/(2\zeta)-\ln(\zeta+\sqrt{\zeta^2-1})-
\label{OP2}\\
&-&(\zeta/2)\arctan\left( 1/\sqrt{\zeta^2-1} \right), \qquad \zeta\geq 1,
\nonumber
\end{eqnarray} 
with $\zeta$ being now a function of a new pair-breaking parameter 
$\zeta_0=\hbar/(\tau_{pb}\Delta_0)$ ranging from zero to the critical value $\zeta_0=0.5$ 
at which $\zeta=\infty$ and $\Delta=0$~\cite{AG,Maki}. 

Inserting Eqs.~(\ref{A}) and (\ref{SA}) for $\hat{s}_A(E)$ 
into Eq.~(\ref{I}) and taking the limit $T\to 0$  
we obtain the Josephson current for an arbitrary $\hat{s}_N(E)$ 
as 
\!
\begin{eqnarray}
&&
I=-\frac{4e}{h}
\int_{0}^{\infty}d\omega
\frac{\partial}{\partial\varphi}
\ln\,
\left\{\right.
1+
\nonumber\\
&&
\alpha^4\,{\rm Det}\,s_{ee}(E)\,{\rm Det}\,s^*_{ee}(-E)-
\nonumber\\
&&
\alpha^2\,\left[r_{11}(E)r^*_{11}(-E)+r_{22}(E)r^*_{22}(-E)+\right.
\label{I0}\\
&&
\left.
\left.
{\rm e}^{-i\varphi}t_{21}(E)t^*_{12}(-E)+ {\rm e}^{i\varphi}t_{12}(E)t^*_{21}(-E)
\right]
\right\}_{E=i\omega.}
\nonumber
\end{eqnarray} 
%

\begin{figure}[t]
\begin{center}
\epsfxsize=0.8\hsize
\epsffile{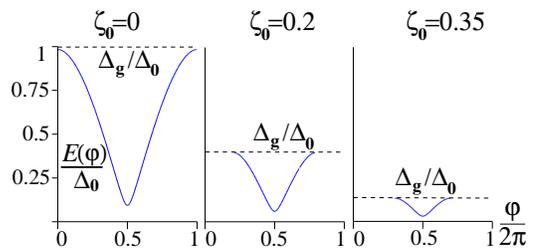}
\end{center}
\caption{Phase dependence of the Andreev bound state for a broad resonant level 
with $\Gamma=15\Delta_0$ close to the Fermi energy ($E_r=0.1\Gamma$);
dashed line shows the normalized gap for a given value of the pair-breaking parameter 
$\zeta_0$.}
\label{E15}
\end{figure}
%

\section{Andreev bound states in a resonant junction}

Let us assume that the $N$ region is a small quantum dot  
and electrons can only tunnel via one of its levels
characterized by its position $E_r$ 
with respect to the Fermi level and broadening $\Gamma$. 
For the simplest Breit-Wigner scattering matrix with 
$r_{11}=r_{22}=(E-E_r)/(E-E_r+i\Gamma)$ and 
$t_{12}=t_{21}=\Gamma/i(E-E_r+i\Gamma)$, 
equation (\ref{I0}) reads
\!
\begin{eqnarray}
&
I=-(2e/h){\cal T}\sin\varphi
\int\limits_{0}^{\infty}d\omega
\times&
\label{I1}\\
&
\left\{ 
u^2\left( 
{\cal R}+{\cal T}\left( 1+ \frac{ \sqrt{1-u^2}-\zeta  }{\Gamma/\Delta} \right)^2
\right)
-1+{\cal T}\sin^2\left(\frac{\varphi}{2}\right)
\right\}^{-1}_{E=i\omega.}&
\nonumber
\end{eqnarray} 
where ${\cal T}=1-{\cal R}=\Gamma^2/(E^2_r+\Gamma^2)$ 
is the Breit-Wigner transmission probability 
at the Fermi level. 
The parameter $\Gamma/\Delta$ accounts for the energy dependence 
of the resonant superconducting tunneling.
In Eq.~(\ref{I1}) the integrand has, in general, poles given by the equation 
\begin{eqnarray}
&
u^2\left( 
{\cal R}+{\cal T}\left( 1+ \frac{ \sqrt{1-u^2}-\zeta  }{\Gamma/\Delta} \right)^2
\right)
=1-{\cal T}\sin^2\left(\frac{\varphi}{2}\right).&
\label{Eq}
\end{eqnarray} 
Along with Eq.~(\ref{u}) they determine  
the energies of the Andreev bound states (ABS) localized 
below the Abrikosov-Gorkov gap $\Delta_{\bf g}=\Delta\left(1-\zeta^{2/3}\right)^{3/2}$.
It is instructive to understand how the pair breaking modifies the ABS spectrum 
since this is reflected on both the current-phase relation $I(\varphi)$ 
and the critical current $I_c\equiv\max\,I(\varphi)$.

We start our analysis with an analytically accessible case 
of an infinitely broad resonant level,
$\Gamma/\Delta\to\infty$, where Eq.~(\ref{Eq}) reduces to 
$u^2=1-{\cal T}\sin^2(\varphi/2)$, yielding the ABS energies $\pm E(\varphi)$ [see, Eq.~(\ref{u})]:
\!
\begin{eqnarray}
E(\varphi)=\Delta\sqrt{1-{\cal T}\sin^2(\varphi/2)}
\left[
1-\frac{\zeta}{\sqrt{{\cal T}}|\sin(\varphi/2)|}
\right].
\label{broad}
\end{eqnarray}
Requiring $E(\varphi)\leq\Delta_{\bf g}$ we find that
the ABS exist in the phase interval where $\sin^2(\varphi/2)\geq\zeta^{2/3}/{\cal T}$ 
and only if $\zeta^{2/3}\leq {\cal T}$. 
The numerical solution of Eqs.~(\ref{u}), (\ref{OP1}) and (\ref{Eq}) 
confirms that the interval of the existence of ABS 
gradually shrinks from $0\leq\varphi\leq 2\pi$
to a narrower one with increasing pair breaking [see, Fig.~\ref{E15}].
Outside this interval the Josephson current is carried by the continuum states 
($E\geq\Delta_{\bf g}$) alone, which is automatically accounted for by Eq.~(\ref{I1}). 
An equation of the same form as Eq.~(\ref{broad}) 
was derived earlier for a nonresonant system 
and by a different method~\cite{Averin}.

\begin{figure}[t]
\begin{center}
\epsfxsize=0.8\hsize
\epsffile{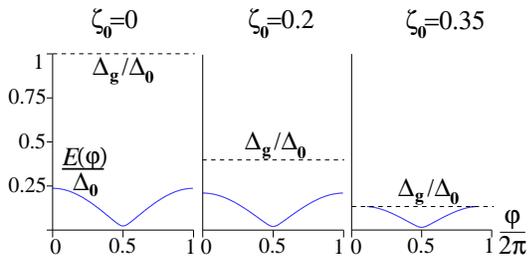}
\end{center}
\caption{
Phase dependence of the Andreev bound state for a narrow resonant level 
with $\Gamma=0.3\Delta_0$ and $E_r=0.1\Gamma$.
}
\label{E03}
\end{figure}

By contrast, the ABS spectrum for a narrow resonant level 
turns out to be much less sensitive to pair breaking. 
Indeed, under condition $\Gamma/\Delta\ll 1-\zeta$ equations~(\ref{u}) and 
(\ref{Eq}) reproduce the known result, 
$E(\varphi)=\sqrt{E^2_r+\Gamma^2}\sqrt{1-{\cal T}\sin^2(\varphi/2)}$ 
[Refs.~\onlinecite{vanHouten,Golubov}].  
In particular, for $E_r\to 0$ the ABS exist within the resonance width, 
$E(\varphi)<\Gamma$
and are separated from the continuum by a gap $\Delta_{\bf g}-\Gamma$. 
Solving Eqs.~(\ref{u}), (\ref{OP1}) and (\ref{Eq}) numerically, 
we find that until this gap closes at a certain value of $\zeta_0$, 
the ABS spectrum remains virtually intact [see, Fig.~\ref{E03}]. 
For bigger $\zeta_0$, the spectrum gets modified in a way similar 
to the previous case [cf., third panels in Figs.~\ref{E15} and~\ref{E03}].
In the case of a very narrow resonance, the characteristic value of $\zeta_0$ is $\approx 0.45$, 
corresponding to $\zeta\approx 1$, 
i.e. to the onset of gapless superconductivity~\cite{AG,Maki}.

\section{Critical current and current-phase relation: Results and Discussion}
 
For numerical evaluation of the Josephson current (\ref{I1}) 
we first put $E=i\omega$ in Eq.~(\ref{u}) and then make the transformation 
$u\to i\nu$, yielding $\omega/\Delta=\nu(1-\zeta/\sqrt{1+\nu^2})$. 
Using this relation, in Eq.~(\ref{I1}) we change to the integration over $\nu$ 
with the Jakobian $d\omega/d\nu=\Delta[1-\zeta/(1+\nu^2)^{3/2}]$:
\!
\begin{eqnarray}
&
I=(2e\Delta/h){\cal T}\sin\varphi
\int\limits_{\nu_0}^{\infty}d\nu\,\left(1-\frac{\zeta}{(1+\nu^2)^{3/2}}\right)
\times&
\label{I2}\\
&
\left\{ 
\nu^2\left( 
{\cal R}+{\cal T}\left( 1+ \frac{ \sqrt{1+\nu^2}-\zeta  }{\Gamma/\Delta} \right)^2
\right)
+1-{\cal T}\sin^2\left(\frac{\varphi}{2}\right)
\right\}^{-1}.&
\nonumber
\end{eqnarray} 
Positiveness of $\omega$ in Eq.~(\ref{I1}) enforces the choice of 
the lower integration limit: $\nu_0=0$ for $\zeta\leq 1$ and $\nu_0=\sqrt{\zeta^2-1}$ 
for $\zeta\geq 1$.

\begin{figure}[t]
\begin{center}
\epsfxsize=0.7\hsize
\epsffile{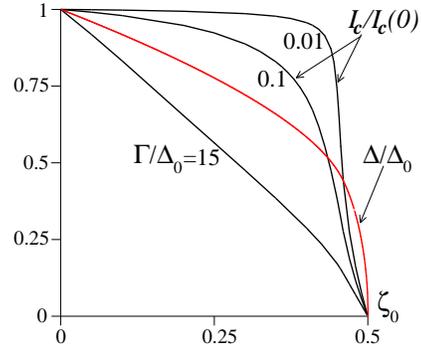}
\end{center}
\caption{On-resonance critical current vs. pair-breaking parameter 
$\zeta_0=\hbar/(\tau_{pb}\Delta_0)$ for different $\Gamma/\Delta_0$. 
The behavior of the normalized order parameter~\cite{AG,Maki}
is shown, for comparison, in red.
}
\label{Iz}
\end{figure}
\begin{figure}[b]
\begin{center}
\epsfxsize=0.7\hsize
\epsffile{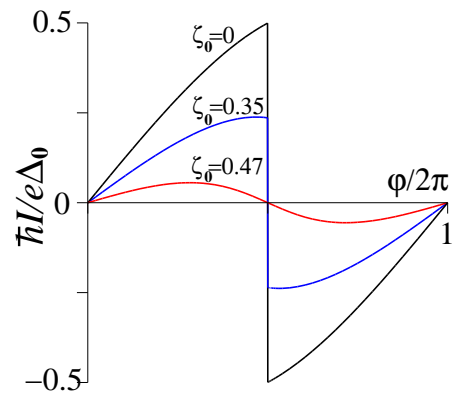}
\end{center}
\caption{
On-resonance current-phase relation for different values 
of the pair-breaking parameter $\zeta_0$ and $\Gamma=\Delta_0$.}
\label{I_f}
\end{figure}

Using Eqs.~(\ref{OP1}), (\ref{OP2}) and (\ref{I2}) we are able to 
analyze the critical current $I_c\equiv\max\,I(\varphi)$ in the whole 
range of the pair-breaking parameter, $0\leq\zeta_0\leq 0.5$ 
[see, Fig.~\ref{Iz}]. 
In line with the discussed behavior of the Andreev bound states,
for a narrow resonance, $\Gamma/\Delta_0\ll 1$, 
the critical current starts to drop significantly only upon entering 
the gapless superconductivity regime $0.45\leq\zeta_0\leq 0.5$.  
On the other hand, for a broad resonance, $\Gamma/\Delta_0\gg 1$,  
the suppression of $I_c$ is almost linear in the whole range. 
We note that in both cases the behavior of $I_c$ 
strongly deviates from that of the bulk order parameter (red curve)~\cite{AG,Maki} 
largely due to the pair-breaking effect on the ABS. 
In practice, the $I_c(\zeta_0)$ dependence can be measured by applying  
a magnetic field [the case where $\zeta_0=(B/B_*)^2$ and 
$B_*=(\Phi_0/\pi d)\sqrt{18\Delta_0/\hbar v_F\ell}$] 
in an experiment similar to Ref.~\onlinecite{Jarillo} 
where a quantum dot, defined in a single-wall carbon nanotube, 
was strongly coupled to the leads with the ratio $\Gamma/\Delta_0\approx 10$.
Carbon nanotube quantum dots with lower $\Gamma/\Delta_0$ values are  
accessible experimentally, too~\cite{Basel,SQUID}.

\begin{figure}[t]
\begin{center}
\epsfxsize=0.95\hsize
\epsffile{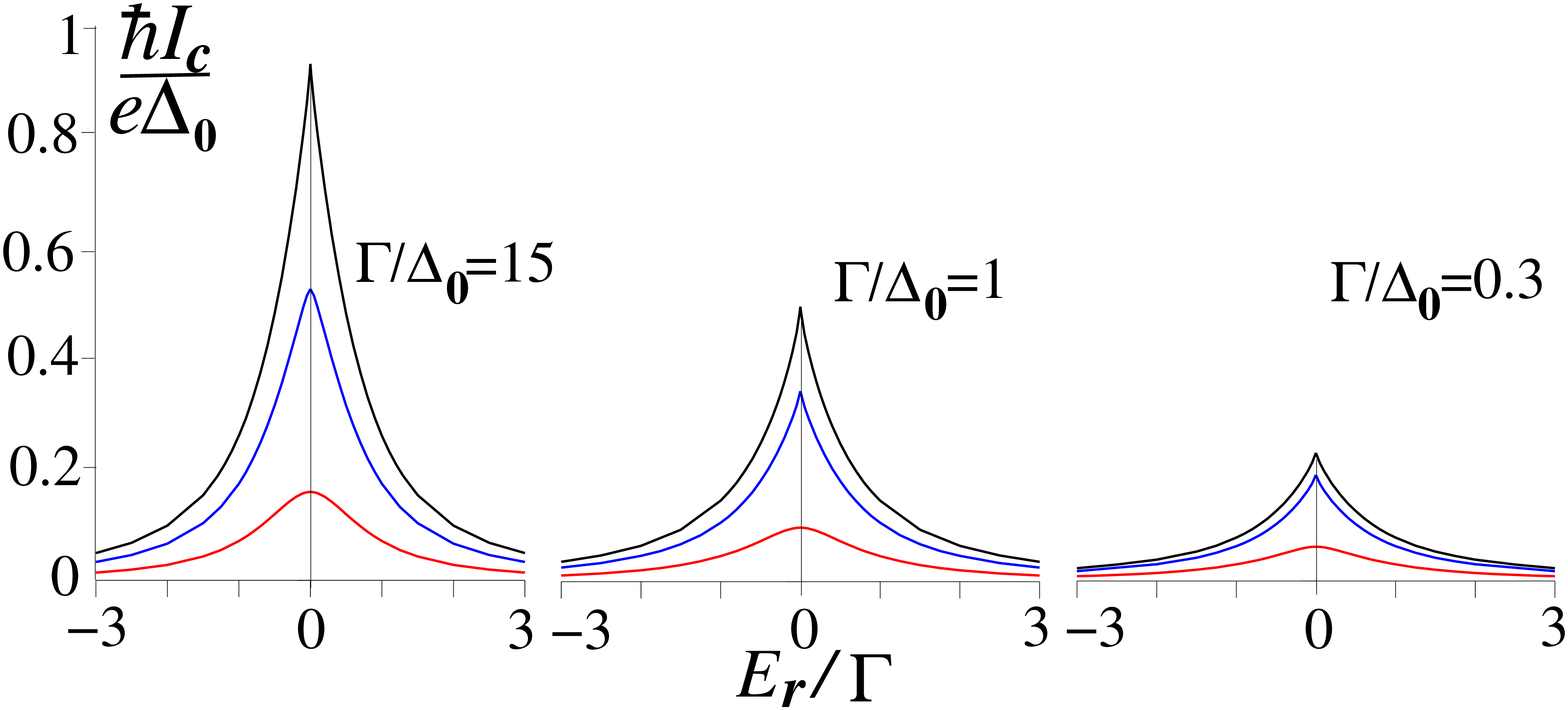}
\end{center}
\caption{Critical current vs. resonant level position: 
$\zeta_0=0$ (black), $\zeta_0=0.25$ (blue) and $\zeta_0=0.45$ (red).}
\label{I_E}
\end{figure}

We also found that the crossover between the gapped and gapless regimes 
is accompanied by a qualitative change in the shape of the Josephson 
current-phase relation $I(\varphi)$ as demonstrated in Fig.~\ref{I_f} 
for the on-resonance case $E_r=0$ and $\Gamma=\Delta_0$. 
The $I(\varphi)$ relation is anharmonic 
as long as the junction with $\Delta_{\bf g}\not=0$ supports the ABS (black and blue curves). 
The vanishing of the ABS upon entering the gapless regime 
leads to a nearly sinusoidal current-phase relation (red curve). 
A closely related effect 
is demonstrated in Fig.~\ref{I_E} showing the modification of the critical current resonance 
lineshape with the increasing pair-breaking strength. 
In the absence of pair breaking it is nonanalytic near $E_r=0$ (black curves) 
reflecting the anharmonic $I(\varphi)$ due to the ABS in a transparent channel~\cite{vanHouten,Beenakker}. 
On approaching the gapless regime this singularity is smeared out 
(red curves), which is accompanied by 
the suppression of the $I_c$ amplitude. 
At finite temperatures $T\ll\Delta/k_B$ 
the pair-breaking-induced smearing of the resonance peak 
will enhance the usual temperature effect.

In conclusion, we have proposed a model describing resonant Josephson tunneling through 
a quantum dot beyond the conventional BCS picture of the superconducting state 
in the leads. It allows for nonperturbative treatment of pair-breaking processes induced 
by a magnetic field or paramagnetic impurities in diffusive superconductors. 
We considered no Coulomb blockade effects, assuming small charging energy in the dot 
$E_C\ll\Delta_0,\Gamma$, which was, for instance, the case in the experiment of Ref.~\onlinecite{Jarillo}. 
Our predictions, however, should be qualitatively correct also 
for weakly coupled dots with $\Gamma\leq E_C\ll\Delta_0$ at least as far as 
the dependence of the ctitical supercurrent 
on the pair-breaking parameter is concerned. 
Indeed, for a narrow resonance the Andreev bound states 
begin to respond to pair breaking only when the gap $\Delta_{\bf g}$ becomes sufficiently small 
[see, Fig.~\ref{E03}] so that 
for a finite $E_C\ll\Delta_0$ one can expect a sharp transition to the resistive state, too,  
similar to that shown in Fig.~\ref{Iz} for $\Gamma/\Delta_0\ll 1$. 

We thank D. Averin, C. Bruder, P. Fulde, A. Golubov, M. Hentschel, T. Novotny, 
V. Ryazanov and C. Strunk for useful discussions.
Financial support by the Deutsche Forschungsgemeinschaft 
(GRK 638 at Regensburg University) is gratefully acknowledged.



\end{document}